# Speed and Reliability of Nanomagnetic Logic Technology

Zheng Gu[1], Mark E. Nowakowski[1], David B. Carlton[2], Ralph Storz[3], Jeongmin Hong[1], Weilun Chao[2], Brian Lambson[4], Patrick Bennett[1], Mohmmad T. Alam[5], Matthew A. Marcus[6], Andrew Doran[6], Anthony Young[6], Andreas Scholl[6], and Jeffrey Bokor[1]

[1]*Department of Electrical Engineering and Computer Sciences, University of California, Berkeley, California 94720, USA*
[2]*Center for X-ray Optics, Lawrence Berkeley National Laboratory, Berkeley, California 94720, USA*
[3]*Thorlabs Inc., 56 Sparta Ave., Newton, NJ 07860, USA*
[4]*iRunway, 2906 Stender Way, Santa Clara, CA 95054, USA*
[5]*Intel Corp., 5200 NE Elam Young Pkwy, Hillsboro, OR 97124, USA*
[6]*Advanced Light Source, Lawrence Berkeley National Laboratory, Berkeley, California 94720, USA*

**Abstract**

Nanomagnetic logic is an energy efficient computing architecture that relies on the dipole field coupling of neighboring magnets to transmit and process binary information. In this architecture, nanomagnet chains act as local interconnects. To assess the merits of this technology, the speed and reliability of magnetic signal transmission along these chains must be experimentally determined. In this work, time-resolved pump-probe x-ray photo-emission electron microscopy is used to observe magnetic signal transmission along a chain of nanomagnets. We resolve successive error-free switching events in a single nanomagnet chain at speeds on the order of 100 ps per nanomagnet, consistent with predictions based on micromagnetic modeling. Errors which disrupt transmission are also observed. We discuss the nature of these errors, and approaches for achieving reliable operation.

Nanomagnetic logic (NML) is a spintronics-based logic architecture which performs basic computational operations with potential for ultralow energy dissipation per operation[1]. In NML, the magnetization of single-domain ferromagnetic thin-film islands, spaced apart by tens of nanometers, are coupled by dipolar fields generated from adjacent islands. The magnetization of individual islands is engineered using shape anisotropy to align either up or down along their magnetic easy axis. When islands are arranged in a line parallel to their magnetic hard axes, the nearest-neighbor magnetic dipolar field coupling imparts a preference for neighboring islands to align anti-parallel, forming a spatial logical inverter. An extended series of these closely spaced nanomagnets, called a chain, transmits binary information from one end to the other sequentially, performing a function similar to integrated circuit (IC) interconnects. An arrangement of three islands around a central island, known as a majority logic gate, produces an output magnetization state based on the majority vote of the three input magnetizations[1].

Energy dissipation calculations for nanomagnets predict less than 1 eV per switching event[1] with no standby leakage power. These energy efficiency predictions (assuming 100 ps switching events[1-3]) are more than an order of magnitude lower than the lowest energy dissipation projections for nanoscale IC transistors[4]. Thus, NML architectures are potential candidates for new energy efficient computing systems that, if fully optimized, could even approach the theoretical minimum amount of energy to perform a single operation set by fundamental thermodynamic limits[5]. The promising potential of this technology has motivated experimental and computational studies of the architecture. In particular, there has been great interest in studying the behavior of chains because they encapsulate much of the essential physics of the architecture. These studies have modeled transmission dynamics[2,6], analyzed the phenomenology of transmission errors[7,8], and demonstrated quasi-static transmission in chains with in-plane[9,10] and out-of-plane[11,12] magnetic easy axes. However, no experimental study thus far has observed the dynamics of magnetic transmission along these chains on nanosecond timescales. In this work, we experimentally demonstrate and observe successful magnetic signal propagation in an NML chain at such timescales and identify and discuss the critical factors which contribute to transmission errors in these systems.

To perform multiple, successive logic operations, the entire NML architecture (chains and majority logic gates) must be re-initialized after each operation. This resetting process is known as clocking and in this work it is performed using an on-chip magnetic field to drive the magnetization of all nanomagnets in a chain to saturation along their magnetic hard axes. This places each magnet in an energetically unstable (null) state which, upon removal of the clock field, is influenced by the dipolar coupling from a neighboring magnet. The time-dependent relaxation from the null state in rectangular or elliptically shaped magnets, however, can also be affected by factors such as thermal fluctuations[2,3, 9,10] and non-ideal magnetic anisotropies[8]. These environmental factors can drive the magnetization of individual islands to spontaneously switch out of sequence, spoiling the dipolar coupling 'cascade' which correctly propagates the input information.

One method to improve the stability of the null state is to fabricate lithographically-defined notches on both ends of each nanomagnet long axis to artificially engineer a metastable potential well along the hard axis by introducing a biaxial anisotropy term[9]. This anisotropy engineering reduces the influence of thermal fluctuations and random lithographic variations that drive spontaneous magnetization relaxation in individual nanomagnets after removal of the clocking field. Error-free transmission along chains with lithographically-enhanced stability has been demonstrated but was driven by quasi-static thermal excitation[10]. For NML to be a viable alternative logic architecture, error-free transmission must occur reliably at very fast time scales. To further investigate the feasibility of this architecture as a practical computing platform, we studied its performance at its natural speed limits. In this work we directly image the data transmission process of NML chains at fast timescales using time-resolved photo-emission electron microscopy (PEEM)[13]. Nanomagnet chains are fabricated on a metal strip, through which current pulses synchronized with x-ray pulses from the Advanced Light Source (ALS) synchrotron at Lawrence Berkeley National Lab generate repeatable clock fields used to perform a stroboscopic "pump-probe" measurement. We observe transmission times on the order of 100 ps per nanomagnet which is consistent with previous predictions[1-3].

Devices are fabricated on silicon wafers coated with a 100 nm thick insulating layer of $SiO_2$. A gold metal strip, 6 µm wide and 158 nm thick is patterned on the surface using photolithography, thermal evaporation and lift-off. A 45 nm thick layer of spin-on aluminum oxide phosphate (AlPO) glass (Inpria Corp.) was deposited on the sample in order to provide a smoother surface on which to form the nanomagnet chains atop the Au strip. Without the AlPO layer, the surface roughness of the Au after patterning is 1.9nm RMS, as measured by atomic force microscopy (AFM). The surface roughness with the AlPO overlayer is 0.35nm. This reduction is important because surface roughness has been found to contribute to irregularities in the shape of the nanomagnets formed on top and thus contributes to transmission errors[14]. Next, chains of 12 nanomagnets bounded by shape-biased inputs, which set the magnetic orientation of the first magnet in each chain,[15] and blocks, which stabilize the final magnet in each chain,[14] are patterned by electron beam lithography. The nanomagnets in each chain are 150 nm wide with 30 nm gaps between them, and vary in height from 300 nm to 500 nm. Notches are included in the magnet shape.[9] The nanomagnets are made by electron beam evaporation and lift-off of 2 nm of titanium, 12 nm of Permalloy (Py), and 2 nm of aluminum as a capping layer. Figure 1a shows a scanning electron micrograph (SEM) of a chain in this pattern. A final 1 nm film of platinum is sputtered over the entire sample surface to reduce surface charging during the PEEM measurement.

Our dynamics experiment is set up as a stroboscopic pump-probe measurement[16,17] where fast current pulses along the strip are synchronized with x-ray pulses from the ALS. We designed and built a customized sample holder to generate the current pulses while isolating the critical electronic components from the PEEM high-voltage electron optics. The requirement in the PEEM instrument to bias the entire sample holder at high-voltage (12-18 kV) and the need for sub-nsec rise and fall time dictated the use of an optical link in place of electrical feedthroughs to trigger the current pulse. The in-situ pulser circuit was built from commercially-available surface mount components, an avalanche photodiode (APD), and a polyimide printed circuit board (PCB) with a silver plated copper conductor. When the APD is irradiated by a short, infrared laser pulse, it produces an electrical pulse that is amplified by two stages of radio frequency amplifiers. Current pulses with a peak amplitude of up to 1 A are delivered into the gold strip by this circuit. A near field calculation using the superposition integral formulation of the Biot-Savart law estimates that the amplified pulses produce an in-plane magnetic field with a peak amplitude of 1000 G at the location of the chains on the surface of the strip. This field is our clock mechanism. The sample and circuit assembly contacts a copper heatsink to dissipate heat from both amplifiers. The probe beam consists of 70 ps x-ray pulses at a repetition rate of 3 MHz. 850 nm pulses with 1-2 nsec duration from a fast diode laser strike the APD at the same repetition rate. The diode laser is mounted outside the PEEM vacuum chamber and the beam is free-space coupled through a window port and focused onto the APD. A pulse delay generator triggered by a synch pulse from the synchrotron RF system synchronizes the x-ray bunches to the laser pulse. Accumulating magnetic contrast images in the PEEM for varying delay allows us to study the time dynamics. Figures 1b and 1c show a schematic of the experiment. To obtain magnetic contrast by x-ray magnetic circular dichroism (XMCD)[18], we tune the x-rays to the Fe $L_3$ absorption edge, expose two images at the same location illuminated by right and left circularly polarized x-rays, respectively, and then compare them by per-pixel numerical division. The images are then adjusted using a median noise filter and linear brightness and contrast stretching calibrated with nearby non-magnetic regions. We imaged 22 chains and each image is averaged over 360 million clock cycles (2 minutes integration time).

Figure 1d shows photo-emission yield from an image of the conducting strip versus time delay. During imaging, Lorentz forces generated by the clock field deflect emitted photoelectrons during a clock pulse, shifting the image and reducing the image intensity. Using this effect we characterize our magnetic field pulse width to be 2 ns and assign zero delay at the peak of the pulse. Figure 2 is a series of XMCD-PEEM images of one chain of nanomagnets (input at the bottom) taken at time delays between 1 and 4 ns. In the image brightness scale, black and white correspond to opposite magnetizations along the horizontal axis. Due to the multiple cycle averaging, the color contrast is diminished when the switching is not fully repeatable on each cycle. This chain exhibits relatively few transmission errors, and prominently demonstrates a 'cascade-like'[9] sequential switching behavior. The red bars in each image are eye-guides which highlight signal transmission. The signal appears to transmit an average of 7 nanomagnets up the chain before encountering a defect caused by an unstable nanomagnet. We can estimate from these images an approximate switching time of just over 100 ps per nanomagnet, which is consistent with predictions made from simulations and calculations[1-3]. This high-speed regime of switching is governed by the Landau-Lifshitz-Gilbert dynamics[3] and is distinct from the typically slower thermally-assisted switching that follows a modified Arrhenius formulation[19]. We also observed nanomagnet switching on longer time scales. Some nanomagnets that showed no magnetic contrast before 4 ns switched by 326 ns. We believe that these nanomagnets remain aligned with their magnetic hard axes for an extended period of time, and eventually switch by thermal activation. Similar thermally assisted switching has been observed on minute and even hour time scales in an earlier work on signal transmission using similar nanomagnet geometries[10].

These observations of NML chain transmission dynamics suggest that stringent requirements for the precision and reproducibility of nanomagnet fabrication are necessary to achieve consistent high-speed transmission. Error-free, fast switching is needed for a practical technology. The thermally activated switching studied in previous work does not meet these requirements due to both slow and temperature-dependent variability in transmission speeds. Fast switching provides reasonable speed but places further constraints on nanomagnet design, because it requires that the potential barrier of the hard-axis energy well induced by the notches be completely lowered by dipole field coupling. Achieving this demands a precisely engineered energy landscape that balances thermal energy and dipole field coupling from the neighboring nanomagnet. A large energy barrier inhibits spontaneous thermal switching, but upon dipole field biasing, a residual potential well may inhibit fast relaxation to the easy axis (Figure 3a,b). Alternatively, a small energy barrier can be easily switched by dipole field coupling (Figure 3c,d) but is also more susceptible to spontaneous thermal switching (Figure 3e). Therefore, to harness the efficient computation potential of NML at fast timescales, careful optimization is required to manage and balance these tradeoffs.

To maximize transmission efficiency, the uniaxial ($K_u$) and biaxial ($K_b$) anisotropy energies of individual nanomagnets in each chain must be precisely engineered within a specific energy range. The limits of this range are directly related to the nanomagnet dimensions[2]. The magnets imaged in Figure 2 were purposefully engineered with $K_u$ = 21.8 kJ/m$^3$ and $K_b$ = 23.8 kJ/m$^3$, which are within the narrow range of error-tolerance as calculated by macrospin simulations (Figure 4a). In this simulation we plot the error-free transmission probability of a chain as a function of the angular misalignment of the central (6th) magnet[8]. Each data point represents 100 simulations at 300 K using a stochastic Landau-Lifshitz-Gilbert evolver. The plot shows that our experimental sample (green) lies within a small tolerance window, between (0.915 – 0.973) $K_u/K_b$, that allows the fabricated chain to accept an angular misalignment of approximately 0.5°. Looking forward, we identify two pathways to improve future tolerances. First, improving lithography will create more consistent biaxial engineering from magnet-to-magnet; less variation will help reduce the number of misalignment errors. Additionally, increasing dipolar coupling between nanomagnets by scaling down their dimensions and using higher magnetization materials will increase the amount of angular misalignment tolerance beyond 0.5 degrees[8]. Identical simulations for 50 nm nanomagnets (spaced 20 nm apart) with approximately identical aspect ratios, $K_u$ = 7.12 kJ/m$^3$, and $K_b$ = 7.47 kJ/m$^3$ show a misalignment tolerance of up to 4°.

For this experiment, we also believe the surface roughness from the Cu wire altered local anisotropies creating larger variations of $K_u$ and $K_b$ from magnet-to-magnet. To determine the role of surface roughness, we compared samples with identical patterns of nanomagnets deposited on 4 surfaces: thermally grown $SiO_2$, 135 nm of AlPO on thermal oxide, 160 nm of copper on thermal oxide, and the AlPO/gold stack described earlier. We reset all samples by an external (non-pulsed) magnetic field of 2200 G for 30 s and then imaged using static PEEM to count the number of errors formed (defined as two consecutive nanomagnets aligned parallel) in a given region on each sample. The copper sample produced 100% more errors than the thermal oxide sample, and the AlPO/gold sample produced 25% more versus the thermal oxide sample. The surface roughness for copper was measured to be 1.6 nm, and compared to the AlPO/gold (0.35 nm) it produced far more errors. We speculate that non-idealities associated with roughness of the substrate layer on which the Ti/Py layers are deposited translate to effective variations in anisotropy of the patterned

Py nanomagnets. This results in a higher probability for defective nanomagnets which leads to an increased likelihood of transmission errors.

In addition to surface roughness, other fabrication-related imperfections have also been identified as a cause of errors[8]. Fabrication defects result in errors involving the defective nanomagnets that repeat in most clock cycles. Figure 3f illustrates how anisotropy asymmetries caused by fabrication defects can lead to repeatable errors. To examine the repeatability of errors on the AlPO/gold sample, we delivered clock pulses through the sample using the same pump circuitry, but performed magnetic force microscopy (MFM) on the sample after each clock cycle. We repeated this process for 5 clock cycles on 40 distinct chains, and then recorded the locations of all errors. Since an error involves a pair of nanomagnets, if either member of the pair becomes part of an error in another clock cycle it is considered a repeated error. Figure 4b shows an example of a chain that contained an error that repeated on all 5 clock cycles. Many such repeating errors occurred throughout the sample, suggesting that particular nanomagnets are defective. It follows that the source of these defects are fabrication-related. The repeatability of these errors is also evident in the strong magnetic contrast that they show during pump-probe measurement.

We have demonstrated that fast precessional signal transmission is possible in an NML chain with nanomagnets that have lithographically-engineered biaxial anisotropy. Time-resolved XMCD-PEEM imaging confirmed the 100 ps switching time predicted for this NML system, which demonstrates a transmission velocity of nearly 2000 m/s. In the future, to reduce systematic errors and improve transmission probability, the uniaxial and biaxial anisotropies must be engineered with tighter lithography tolerances to improve consistency between individual magnets, smoother surfaces capable of clocking should be examined, and nanomagnet dimensions should be reduced to induce stronger dipolar field coupling to improve error margins.

**Acknowledgements**

We gratefully acknowledge support from the Western Institute of Nanoelectronics, DARPA, the NSF Center for Energy Efficient Electronics Science, and the Office of Science of the US Department of Energy. Work at the Advanced Light Source, Center for X-ray Optics, and the Molecular Foundry at Lawrence Berkeley National Laboratory is supported by the Director, Office of Science, Office of Basic Energy Sciences, US Department of Energy under contract number DE-AC02-05CH11231. In addition, we acknowledge the Marvell Nanofabrication Laboratory for the cleanroom and machine shop facilities.

**Figures**

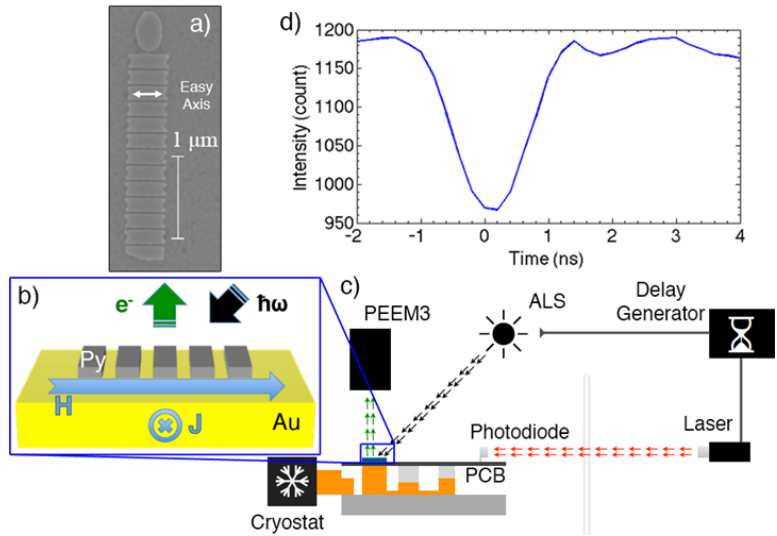

**Figure 1:** a) SEM of a sample chain of permalloy nanomagnets on top of an AlPO/Au stack. The easy axis of an individual nanomagnet is indicated. b) Micro-scale illustration of experimental geometry showing a chain of permalloy (Py) nanomagnets fabricated on top of a Au wire. A current (J) generates the clock field (H) which resets the nanomagnets along their hard axes. This process is observed with x-ray illumination (ℏω) which generates photo-emitted electrons which are captured by electron optics. c) Macro-scale diagram of experiment geometry showing PCB (side view) with attached sample (blue) and APD in relation to experiment equipment. d) Photo-electron intensity count for an image containing the Au wire vs. the time delay.

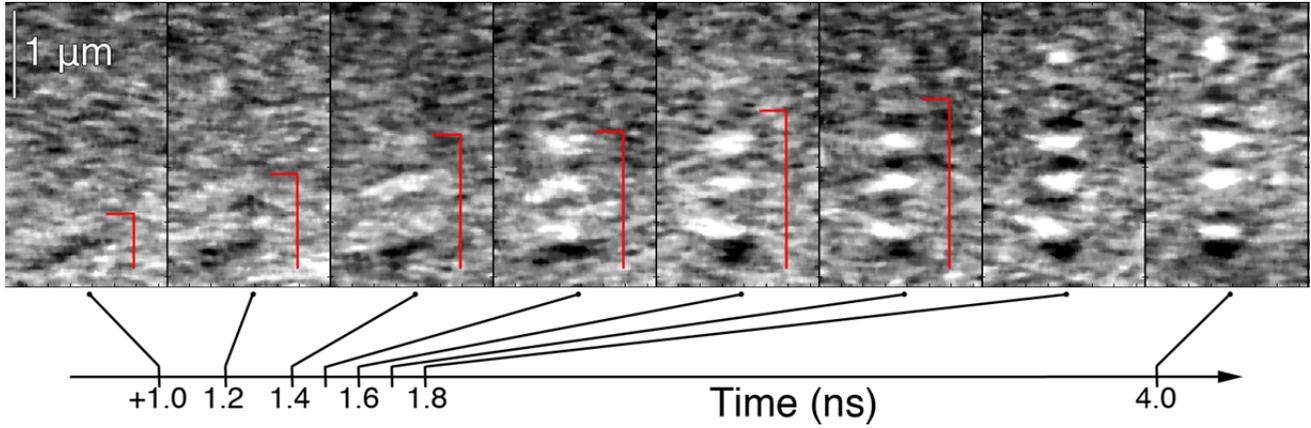

**Figure 2:** XMCD-PEEM images of a nanomagnet chain at various time delays from 1 to 4 ns. The chain demonstrates a 'cascade-like' behavior over time and exemplifies fast bit transmission. Transmission starts at the bottom input magnet. Black and white represent opposite magnetization directions, and the red bars indicate our interpretation of the distance transmitted.

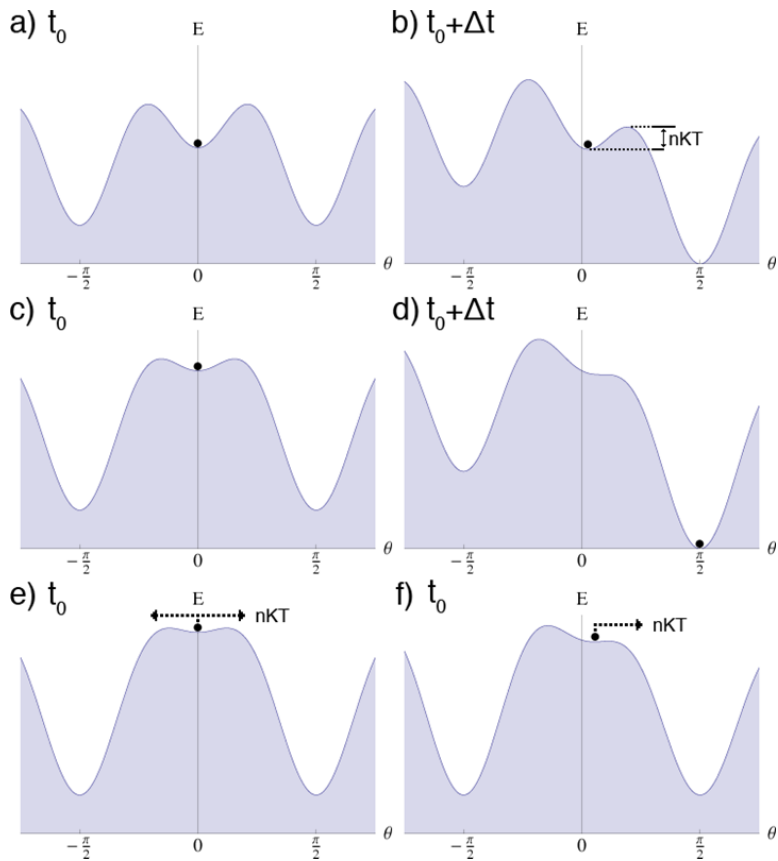

**Figure 3:** Energy vs. angle diagrams (with respect to hard axis) of an individual nanomagnet with lithographically-controlled anisotropy engineering immediately after removing the clock field ($t_0$) then after time has elapsed ($t_0+\Delta t$) to allow the magnetization to equilibrate after precessional switching. The black dots indicate the resulting magnetization directions in the absence of thermal energy. An energy well too deep (a) leads to a residual energy barrier that prevents precessional switching (b). A thermal energy, nKT, is required to switch the nanomagnet. A shallow energy well (c) increases the probability for successful precessional switching (d). However if the well is too shallow it will be more sensitive to errant thermal switching (e). Likewise fabrication defects may cant the metastable state (f) which also increases the likelihood of spontaneous thermal switching.

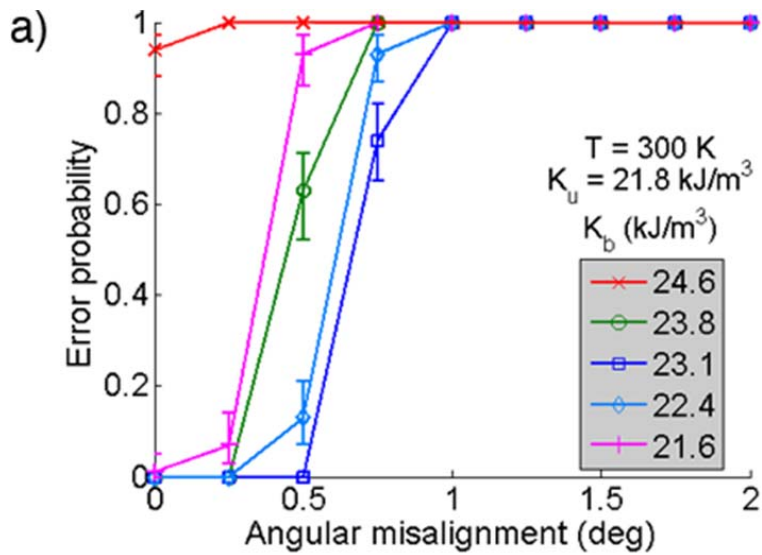

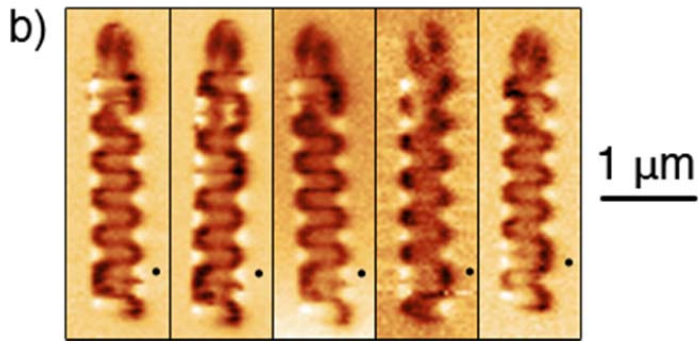

**Figure 4:** a) Macrospin simulations of a chain of 12 nanomagnets at 300 K where the 6th magnet is rotated in-plane by a small angle. The probability of at least 1 error in the chain is plotted after 100 simulations per data point. Extracted values of uniaxial and biaxial anisotropy are used (green) and compared with the cases for small changes in biaxial anisotropy energy density ($K_b$). Only a narrow range of values allows error-free transmission with an angular misalignment tolerance of up to 0.5°. b) MFM images of 5 clock cycles of a chain with a repeated error indicated at the position of the black dot. Dark and light regions on each nanomagnet represent opposite magnetic poles.